# Nonlinear Quasi-Synchronous Multi User Chirp Spread Spectrum Signaling

Nozhan Hosseini, *Member, IEEE*, David W. Matolak, *Senior Member, IEEE*

*Abstract*—**Multi user orthogonal chirp spread spectrum (OCSS) can improve the spectral inefficiency of chirp spread spectrum (CSS) but is only feasible with perfect synchronism and without any channel dispersion. Asynchronism, channel dispersion, or unexpectedly large Doppler shifts can cause multiple access interference (MAI), which degrades performance. Conditions with small timing offsets we term "quasi-synchronous" (QS). In this paper, we propose two new sets of nonlinear chirps to improve CSS system performance in QS conditions. We analytically and numerically evaluate cross-correlation distributions. We also derive the bit error probability for Binary CSS analytically and validate our theoretical result with both numerical and simulation results; our error probability expression is applicable to *any* binary time-frequency (TF) chirp waveform. Finally, we show that in QS conditions our two new nonlinear chirp designs outperform the classical linear chirp and all existing nonlinear chirps from the literature. To complete our analysis, we demonstrate that our nonlinear CSS designs outperform existing chirps in two realistic (empirically modeled) dispersive air to ground channels.**

I. INTRODUCTION

Many wireless communication systems will need to accommodate a larger number of users in the future. One application in particular in which this is critical is low data rate, long range communication links with very large numbers of nodes, such as the internet of things (IoT), internet of flying things (IoFT), etc. These systems demand advanced multi-access techniques with minimal multiple access interference (MAI). They should also be robust to multiple impairments, including multipath channel distortion, Doppler spreading, and interference.

Chirp waveforms [1] can satisfy most of these requirements, and in addition have other attractive features, such as low peak to average power ratio (PAPR) and narrowband interference resilience.

This work was partially supported by NASA, under award number NNX17AJ94A.
The authors are with the Department of Electrical Engineering, University of South Carolina, Columbia, SC (email: nozhan@email.sc.edu, matolak@cec.sc.edu).

Hence chirps—a form of frequency modulation—are promising candidates for many such applications. Chirps are specified in the IEEE 802.15.4a standard as chirp spread spectrum (CSS) [2].

These time frequency (TF) waveforms have several useful properties including energy efficiency, and if wideband enough, robustness to interference, non-linear distortion, multipath fading, and eavesdropping. They can also be used for high resolution ranging and channel estimation. Underwater acoustic wireless communication systems [3], [4] can also use chirp waveforms to advantage in the presence of very rapid underwater acoustic channel fading. The "long-range" (LoRa) technology developed for IoT applications uses a proprietary CSS modulation scheme that aims to provide wide-area, low power and low cost IoT communications [5], [6].

In the literature, different chirp waveforms have been categorized: linear, various types of nonlinear, amplitude variant as well as constant amplitude forms. Modulation can be accomplished in several ways, one of the simplest being binary chirps that sweep either up or down in frequency over a bit period. Chirps can of course be used in on-off signaling or as basic waveforms for frequency shift keying (FSK). Higher-order modulation can be attained with chirps in a number of ways, e.g., by using multiple sub-bands, different start/stop frequencies (somewhat akin to pulse position modulation, PPM), and via distinct chirp waveforms within a given band.

A disadvantage of CSS signaling is spectral inefficiency. This can be addressed by accommodating multiple users with a set of properly designed chirps in the available bandwidth. For multiple access, a set of chirp signals is required, and all waveforms in this set would ideally be orthogonal, without MAI. Achieving orthogonality is easy enough with synchronized waveforms [7]-[10], but in many practical cases, e.g., with mobile platforms, attaining and maintaining synchronism is challenging. Waveforms designed to be orthogonal when synchronized typically exhibit large inter-signal cross correlation when asynchronous [11], [12], and this of course induces MAI and degrades performance. Thus finding a set of waveforms that achieves low cross correlation among signals when asynchronous is desirable. In general this is very difficult in practice, so researchers have focused on quasi-synchronous (QS) conditions. This refers to the case when synchronization is approximate, typically limited to some small fraction of a symbol time $T$. Such approximate synchronization allows for less precise timing control in mobile applications. Example waveforms for this type of application are those based on the zero autocorrelation zone (ZACZ) set of sequences [13].

In this paper, we explore the analysis and design of CSS waveforms for quasi-synchronous operation in multiple access systems using both linear and nonlinear TF functions. We quantify MAI and error probability analytically in asynchronous conditions, and provide two new nonlinear chirp designs that yield better performance under modest asynchronism (i.e., QS).

The remainder of this paper is organized as follows: Section II provides a brief literature review on chirp signaling and CSS, and in Section III we introduce linear and two proposed nonlinear chirp designs. Section IV describes the quasi-synchronous condition and provides analytical and numerical cross correlation results for linear and any nonlinear chirp signal sets; we compare cross correlation values for a full range of delays. Section V addresses analytical binary CSS (BCSS) bit error ratio (BER) performance evaluation and simulation validation. In Section VI, we evaluate QS performance of our proposed waveforms in comparison to linear and other chirp waveforms in the literature, and over an empirical air-ground channel model. Section VII concludes the paper.

## II. LITERATURE REVIEW

The literature on the general use of chirps is fairly extensive (radar, channel modeling, etc.), so we only provide highlights. We focus primarily on communication aspects.

The chirp technique proposal made by S. Darlington in 1947 was related to waveguide transmission for pulsed radar systems with long range performance and high range resolution [14]. B. M Oliver first used "chirp" in his memorandum entitled "not with a Bang, but a Chirp," and 6 years later, acoustic chirp devices were developed at Bell Labs. Hardware constraints were a limiting factor for their development. In [15], the authors described an experimental communication system employing chirp modulation in the HF band for air-ground communication.

In [7], the authors proposed an orthogonal linear amplitude-variant chirp modulation scheme where each user employs a unique frequency modulated chirp rate. The scheme defines orthogonal linear chirps with different chirp rates or TF slopes. To satisfy orthogonality with their design, they impose amplitude variation ($\sim\sqrt{t}$), and hence this scheme does not retain the desirable constant envelope property of conventional chirps. This approach showed improvement in multi user system BER performance in multipath fading channels when compared to FSK frequency hopping code division multiple access (FH-CDMA) schemes. Their analysis and evaluation was based on a perfectly synchronized condition.

The authors of [8] used a set of orthogonal linear chirped waveforms based on the Fresnel transform and its convolution theorem to design an orthogonal chirp division multiplexing (OCDM) system. They compared this to orthogonal frequency division multiplexing (OFDM) and showed that their OCDM system outperformed the conventional OFDM system by exhibiting greater resilience to inter symbol interference when the OFDM system had an insufficient guard interval. Compared to OFDM the OCDM scheme had identical PAPR performance and only slightly higher complexity. Discrete Fourier transform-precoded-OFDM (DFT-P-OFDM) outperformed OCDM in terms of PAPR and had identical BER performance. In this work, the authors also assumed perfect synchronization between all transmitters and receivers.

In [9], the authors presented their orthogonal quadratic and exponential non-linear chirp designs. Users are assigned unique chirp rates that vary either quadratically or exponentially versus time (yielding different signal bandwidths among users). These designs also required amplitude variation to maintain orthogonality. A similar approach was followed in [10] for nonlinear trigonometric and hyperbolic CSS waveforms, again assuming full synchronization.

The authors in [16] presented another set of orthogonal chirps by exploiting the advantages of the fractional Fourier transform (FrFT) adopted from [17]. They claimed that the proposed method has lower MAI than the conventional method in [17] and should yield better system performance. Their signal amplitude is constant over the chirp duration, but again, a fully synchronous system was assumed. The authors in [18] proposed an iterative receiver to improve BER performance in frequency-selective fading channels and opened the possibility of space-time coding multiple input multiple output (MIMO) schemes for orthogonal code division chirps (OCDM). Finally, in [19] and [20] we discussed the implementation of a low complexity transceiver based on discrete Fourier transform spread orthogonal frequency division multiplexing (DFT-s-OFDM). In this work, we provided insight into how chirp waveforms for radar and communication can be synthesized without major modifications to the physical layer of today's OFDM based wireless communication systems. In [21] we investigated air to ground channel fading effects on the BER performance of CSS systems. Specifically, we simulated performance in some "canonical" Ricean fading channels, and over realistic aeronautical channels based on extensive measurements.

Our approach for CSS here enforces constant signal envelope and nearly equal signal bandwidths for all users. Primarily, we relax the perfect synchronization constraint and find designs that can yield better multi-user performance when quasi-synchronous. To the best of our knowledge, this is the first appearance of chirp schemes designed for practical QS operation. The main contributions of this paper can be listed as follows:

- We improve the spectral inefficiency of chirp spread spectrum (CSS) in asynchronous or quasi-synchronous conditions, via introduction of two new nonlinear chirp signals sets.

- We provide analytical and numerical results for cross-correlations for linear and nonlinear CSS systems.

- We derive the bit error probability for binary CSS for any TF waveforms, for arbitrary received signal energies, and validate our theoretical result with both numerical and simulation results.

- We show that in QS conditions our two new nonlinear chirp designs outperform the classical linear chirp and all existing nonlinear chirps from the literature, on the additive white Gaussian noise channel.

### III. Chirp Signal Designs

#### A. Synchronized Linear Chirp Signaling

In this paper, the core formula for generating frequency-modulated (chirp) waveforms is adopted from the kernel Fresnel transform theorem method, related to the Talbot effect, where the discrete Fresnel transform (DFnT) provides the coefficients of the optical field of an image, first observed by Talbot [22]; this is discussed in lightwave and optical communication applications [8]. Using the continuous Fresnel transform provided in [23] and expressed in the form of convolution as noted in [24], we obtain the formula to generate orthogonal linear "up-chirps" (low to high frequency) and "down chirps" (high to low frequency) with symbol duration $T$. In complex baseband form, the $m$th linear chirp waveform can be written as,

$$s_m(t) = e^{\frac{j\pi N}{T^2}\left(t+\frac{mT}{N}\right)^2}, \qquad 0 \leq t < T \qquad (1)$$

where N is the desired number of orthogonal (up-)chirp waveforms, $m \in \{0, 1, \ldots N-1\}$ is the user index, and T is the duration of the chirp waveform. The total bandwidth B that a set of N users occupies is $B=2N/T$, and each user signal occupies the same bandwidth. When perfectly synchronized, the waveforms in (1) are orthogonal. A completely analogous construction can be made with "downchirps" by using a negative sign on the $mT/N$ term of the exponent of (1). In this paper we consider only upchirps, but all assumptions and results are analogous for downchirps. The instantaneous frequency of the signal in (1) can be written as

$$v_m(t) = \frac{1}{2\pi}\frac{d}{dt}\left[\frac{\pi N}{T^2}\left(t^2 + \frac{2mT}{N}t + \frac{m^2 T^2}{N^2}\right)\right] = \frac{N}{T^2}t + \frac{m}{T}. \tag{2}$$

*B. Synchronized Sinusoidal Chirp Signaling*

Non-linear chirp waveforms can easily be generated with arbitrary shapes in the time-frequency plane. The most well-known examples are exponential, quadratic, and sawtooth [9], [10]. Here we propose a mathematical derivation for generating two specific nonlinear chirp waveforms with no amplitude variation (with the aim of keeping PAPR low). A nonlinear phase function Ψ(t) is employed as in (3),

$$s_{m_{NL}}(t) = e^{\frac{j\pi N}{T^2}\left(\left(t+\frac{mT}{N}\right)^2 + \Psi(t)\right)}, 0 \leq t < T. \tag{3}$$

This phase function can modify the instantaneous frequency to any desired nonlinear TF shape. One can find the chirp signal's time-frequency shape via the time derivative $\frac{\Psi'(t)}{2\pi}$ to find instantaneous frequency versus time. We propose two non-linear chirp signal sets which, qualitatively speaking, have more "spacing" between each signal's time-frequency trace. This approach aims to fully use the available time-frequency "space" for signals in a set, and increase resilience to timing offsets for the practical QS case.

Case one uses a sinusoidal function for Ψ(t), with signal waveforms given by,

$$s_{m_{sin}}(t) = e^{\frac{j\pi N}{T^2}\left(\left(t+\frac{mT}{N}\right)^2 + \frac{\alpha t}{2\pi f_0}\sin(2\pi f_0 t)\right)}, 0 \leq t < T \quad (4)$$

where $\alpha$ and $f_o$ are selectable constants that can produce different time/frequency shapes. The instantaneous frequency can be shown to be,

$$v_{m_{sin}}(t) = \frac{N}{T^2}t + \frac{m}{T} + \frac{\alpha N}{4\pi f_c T^2}\sin(2\pi f_0 t) + \frac{\alpha t N}{2T^2}\cos(2\pi f_0 t). \quad (5)$$

We selected values for $\alpha$ and $f_0$ as $\frac{(2m-N)}{2N}$ and $\frac{1}{\pi}$, respectively, as these qualitatively produce a larger "area coverage" in the TF plane than the linear set of signals. An example is plotted in Fig.1.

*C. Synchronized Quartic Chirp Signaling*

In order to further increase spacing between each signal's time/frequency trace, we constructed another nonlinear signal set with the following instantaneous frequency:

$$v_{m_{quartic}}(t) = \begin{cases} \frac{N}{T^2}t + \frac{m}{T} + \beta[t((2t-T)^2 - 2T^2)] & m < \frac{N}{2} \\ \frac{N}{T^2}t + \frac{m}{T} + \beta[(-t+T)((2t-T)^2 - 2T^2)] & m > \frac{N}{2} \end{cases} \quad (6)$$

The corresponding phase functions are,

$$\Psi_{quartic}(t) = \begin{cases} 2\pi\beta\left(t^4 - \frac{4}{3}t^3 T - \frac{1}{2}t^2 T^2\right) & m < \frac{N}{2} \\ 2\pi\beta\left(-t^4 + \frac{8}{3}t^3 T - \frac{3}{2}t^2 T^2 - tT^3\right) & m > \frac{N}{2} \end{cases} \quad (7)$$

where $\beta$ was chosen as $\frac{N-2m}{2NT^2}$. This design yields a larger time/frequency coverage than the linear and sinusoidal nonlinear case.

TF plots of both nonlinear waveforms using (4) and (7) are shown in Fig. 1. Note that not all *N* waveforms are shown: specifically, only the two lowest and highest frequency signals are plotted to bound each signal type's area. The nonlinear cases clearly occupy larger total areas in the TF plane. As Fig. 1 depicts, the sinusoidal Case 1 signal set occupies a slightly larger TF area than the linear set but keeps the same starting and ending frequency and the same total bandwidth. The quartic Case 2 covers the largest area, with different starting and ending frequencies, but the same total bandwidth.

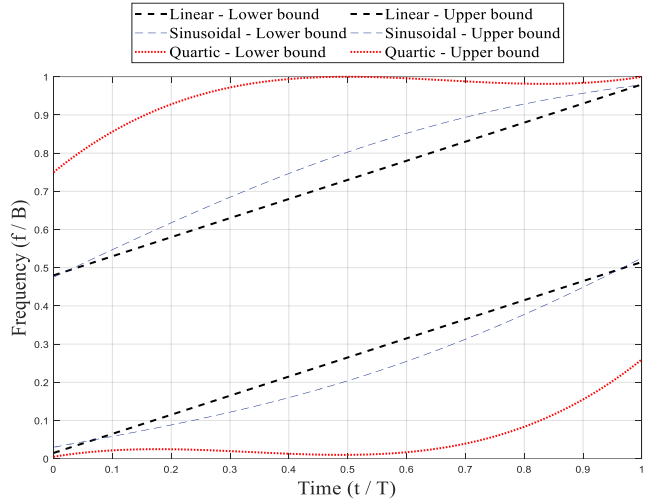

Fig. 1. Time-frequency domain signal representation example showing upper-most and lower-most frequency signals for linear and two nonlinear signal sets.

IV. Quasi-synchronous Transmission

Many modern communication systems have been developed assuming quasi-synchronous conditions, where clocks of different user terminals (or, nodes) are not perfectly synchronized, but are "close" to synchronized. Their mean clock frequencies may be essentially identical, but drift and jitter cause clocks to deviate from this mean over the long and short terms. This asynchronism is usually bounded (a small portion of a symbol duration $\delta T$) in many communication systems. Asynchronism also of course arises from channel effects, primarily propagation delay. Delays are typically modeled as random for all these causes.

For the chirp waveforms of (1), a delayed chirp signal can be written as,

$$s_m(t - \varepsilon_m) = e^{\frac{j\pi N}{T^2}\left[\left((t-\varepsilon_m)+\frac{mT}{N}\right)^2\right]}, \qquad \varepsilon_m \leq t < T + \varepsilon_m \tag{8}$$

where $\varepsilon_m$ is the delay associated with clock drift or uncompensated propagation delay for user *m*. Generally, these delays have value limited between 0 to *T* since other than packet transmission boundaries, effects of asynchronism recur over subsequent symbols (we assume delays are essentially constant over packet durations, and a given user signal uses the same chirp type for each symbols). A time/frequency domain representation of the set of quasi-synchronous signals of the form of (8) for only one asynchronous user (*m=2*) is depicted in Fig. 2. We note that for certain values of timing offset $\varepsilon_2$, the non-synchronized TF signal can overlap another signal in the set nearly completely over a part of a symbol, and this yields relatively large MAI.

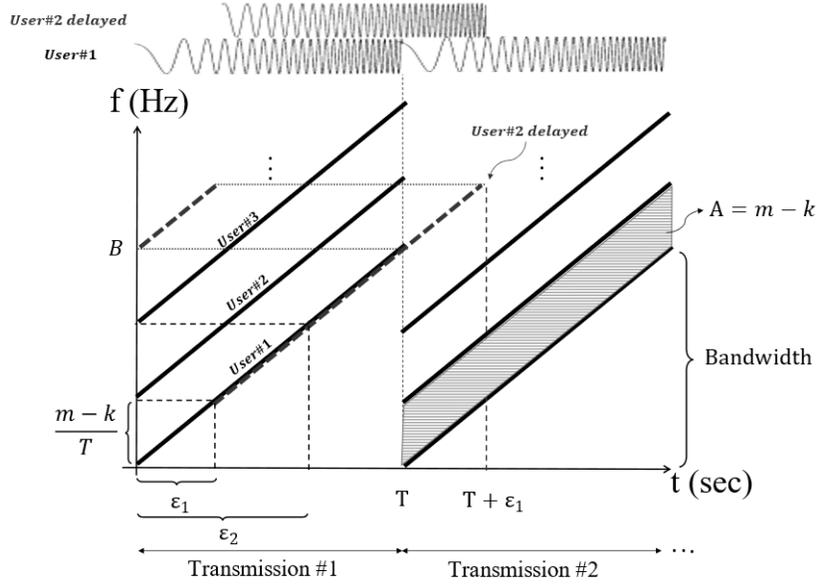

Fig. 2. Time/frequency domain representation of quasi-synchronous transmission of linear chirps with one asynchronous user.

Multiple access interference (MAI) is quantified by the cross correlation between signals of the form of (1) and (8). Computing the cross correlation values requires an integration, which can be written as follows:

$$\rho_{mk}(\varepsilon_k) = \frac{1}{\sqrt{E_m E_k}} \int_0^T s_m(t) s_k^*(t - \varepsilon_k) dt \tag{9}$$

where for the linear chirp case we have,

$$\int_{\varepsilon_k}^{T} e^{\frac{j\pi N}{T^2}\left(t+\frac{mT}{N}\right)^2} \cdot e^{-\frac{j\pi N}{T^2}\left[\left((t-\varepsilon_k)+\frac{kT}{N}\right)^2\right]} + \int_{0}^{\varepsilon_k} e^{\frac{j\pi N}{T^2}\left(t+\frac{mT}{N}\right)^2} \cdot e^{-\frac{j\pi N}{T^2}\left[\left((t+(T-\varepsilon_k))+\frac{kT}{N}\right)^2\right]} \qquad (10)$$

where again $\varepsilon_k$ is the timing offset of user $k$, and we have used the unit-energy of each waveform.

By dividing the integral into two parts as indicated in Fig. 2, this integral has a closed form solution for any arbitrary offset $\varepsilon_k$, and via Euler's identity and l'Hopital's rule, we can find,

$$\rho_{mk}(\varepsilon_k) = \begin{cases} T & (i) \\ \frac{iT^2}{2\pi(kT-mT-\varepsilon_k N)} \left[ \exp\left(\frac{-i\pi(kT-mT-\varepsilon_k N)(kT+mT-\varepsilon_k N+2NT)}{NT^2}\right) - \exp\left(\frac{-i\pi(k^2 T^2-(mT-\varepsilon_k N)^2)}{NT^2}\right) \right] + \\ \frac{iT^2}{2\pi(kT-mT+(T-\varepsilon_k)N)} \left[ \exp\left(\frac{-i\pi(kT-mT+(T-\varepsilon_k)N)(kT+mT+(T-\varepsilon_k)N+2N\varepsilon_k)}{NT^2}\right) - \exp\left(\frac{-i\pi((kT+(T-\varepsilon_k)N)^2-m^2 T^2)}{NT^2}\right) \right] & (ii) \end{cases} \qquad (11)$$

where (i) denotes $\varepsilon_k = (k-m)T/N$ and (ii) denotes otherwise. This expression has the smallest value (0) when $\varepsilon=0$ or $\varepsilon=T$. Correlation is of course one when $m=k$ and $\varepsilon=0$.

The integral for nonlinear chirps has no closed form solution in general, yet for arbitrary nonlinear chirp waveforms one can obtain a very good approximation by modeling any nonlinear TF trajectory as a set of $\mathcal{M}$ linear segments of very small duration. We do not address the mathematical intricacies here, but as $\mathcal{M}$ gets large, for continuous TF functions our approximation should converge to the exact cross correlation integral result. The total cross correlation for any two nonlinear TF functions (integral of (9)) is then the summation of the $\mathcal{M}$ small segments of linear cross correlations. Figure 3 illustrates the approximation method, where for each small segment, a specific linear equation is used to approximate the TF function. Naturally, as the segment length $\Delta T$ decreases (and $\mathcal{M}$ increases), the approximation improves.

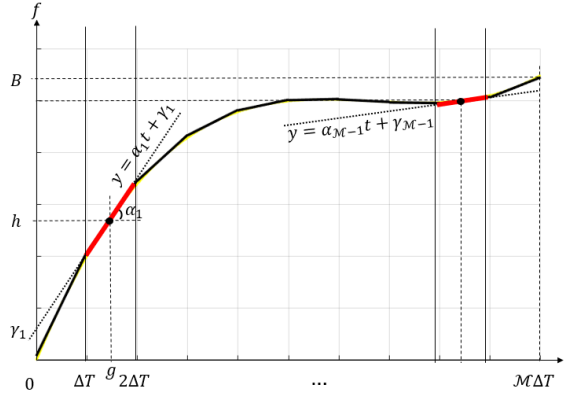

Fig. 3. Method used to find cross correlation for nonlinear TF shapes.

Specifically, in Fig. 3, we show a symbol duration divided into $\mathcal{M}$ equal segments. By taking the derivative of the nonlinear TF function at each segment of width $\Delta T$ we can find the slope ($\alpha$) and intercept ($\gamma$) of each line segment ($y = \alpha t + \gamma$) where $\alpha = v'(g)$, $\gamma = -\alpha g + h$, and $g = \varsigma \Delta T + \frac{\Delta T}{2} = \left(\varsigma + \frac{1}{2}\right)\Delta T$. The variables $h$ and $g$ are constant frequency and time values, respectively, of the center of the line segment, $\varsigma$ is the index on $\Delta T$ and $v$ is the instantaneous frequency as defined previously in (2) and (5). We can then write the nonlinear chirp signal cross correlation for the $m$th and $k$th signals as,

$$\rho_{mk_{NL}}(\varepsilon_k) = \frac{1}{\mathcal{M}} \sum_{\varsigma=0}^{\mathcal{M}-1} \int_{\varsigma \Delta T}^{(\varsigma+1)\Delta T} y_{m\varsigma}(t) y_{k\varsigma}^*(t - \varepsilon_k) dt \tag{12}$$

where $y_{m\varsigma}(t)$ and $y_{k\varsigma}(t)$ can be written as,

$$y_{m\varsigma}(t) \cong e^{j2\pi\left(\int y_{m\varsigma}(t) + \varphi_{m\varsigma}\right)} = e^{j2\pi\left(\frac{\alpha_{m\varsigma}}{2}t^2 + \gamma_{m\varsigma}t + \varphi_{m\varsigma}\right)}$$
$$y_{k\varsigma}^*(t - \varepsilon_k) \cong e^{-j2\pi\left(\int y_{k\varsigma}(t-\varepsilon_k) + \varphi_{k\varsigma}\right)} = e^{-j2\pi\left(\frac{\alpha_{k\varsigma}}{2}t^2 + \gamma_{k\varsigma}t + \varphi_{k\varsigma}\right)} \tag{13}$$

where $\varphi_{m\varsigma}$ and $\varphi_{k\varsigma}$ are the phases of the $\varsigma th$ segment, and $\varphi_{m\varsigma}$ can be written as,

$$\varphi_{m\varsigma} = \frac{\alpha_{m(\varsigma-1)} - \alpha_{m\varsigma}}{2}\varsigma\Delta T^2 + \left(\gamma_{m(\varsigma-1)} - \gamma_{m\varsigma}\right)\varsigma\Delta T, \tag{14}$$

with $\varphi_{k\varsigma}$ defined analogously. Therefore, using (13) and (14) in (12), and making use of the online Wolfram Alpha intelligence computational integral engine [25] the cross correlation of the two nonlinear chirps can be approximated as,

$$\rho_{mk_{NL}}(\varepsilon_k) \cong \sum_{\varsigma=0}^{M-1} -\frac{(-1)^{\frac{3}{4}} \exp\left(i\pi\left(2(\varphi_{m\varsigma} - \varphi_{k\varsigma}) - \frac{(\gamma_{m\varsigma} - \gamma_{k\varsigma})^2}{\alpha_{m\varsigma} - \alpha_{k\varsigma}}\right)\right)}{2\sqrt{(\alpha_{m\varsigma} - \alpha_{k\varsigma})}}$$
$$\times \left[\text{erfi}\left(\frac{\sqrt[4]{-\pi^2}\left((\alpha_{m\varsigma} - \alpha_{k\varsigma})(\varsigma+1)\Delta T + \gamma_{m\varsigma} - \gamma_{k\varsigma}\right)}{\sqrt{\alpha_{m\varsigma} - \alpha_{k\varsigma}}}\right) - \text{erfi}\left(\frac{\sqrt[4]{-\pi^2}\left((\alpha_{m\varsigma} - \alpha_{k\varsigma})(\varsigma)\Delta T + \gamma_{m\varsigma} - \gamma_{k\varsigma}\right)}{\sqrt{\alpha_{m\varsigma} - \alpha_{k\varsigma}}}\right)\right] \tag{15}$$

where $erfi(z)$ is the imaginary error function, defined by $erfi(z) = -ierf(iz)$ where $z$ is a real number. We show an example result in Fig. 4 for cross correlation between two nonlinear user signals for delay of $\epsilon = 0.2T$ using this method. In this particular signal set, the actual TF functions change as the total number of users $N$ changes; hence we show $\rho_{25}$ for two values of $N$. We can see as the number of segments per symbols used in the integral approximation increases, the approximation gets closer to the cross correlation value computed via direct numerical integration of (9).

In general mentioned delays can be well modeled as random, and we can assess the quality of any chirp signal set statistically, by considering cross correlation to be conditioned upon delay, then averaging that over the probability density function of delay. An example set of *mean* correlations is shown in Fig. 5. This figure shows results for a set of *N*=25 linear chirps, and for two sets of *N*=25 nonlinear chirps, both sinusoidal and quartic.

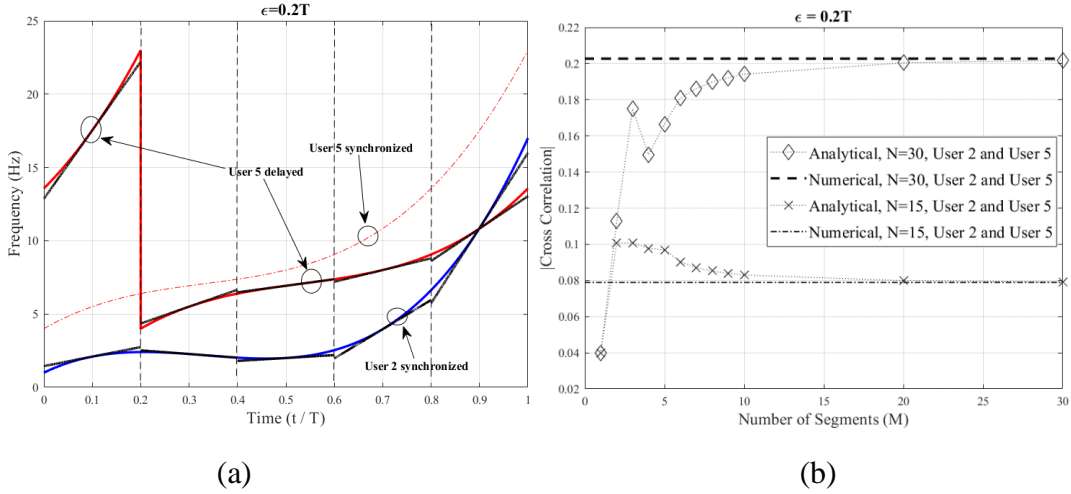

(a)  (b)

Fig. 4. Approximation of nonlinear chirp by linear segments: (a) TF traces for two signals in a set of $N$=15 users, (b) Cross correlation $\rho_{25}$ vs. number of segments ($\mathcal{M}$) for user two and five when in sets of $N$=15 and $N$=30 users.

Here we found cross correlation between each pair of two users in a set with relative delay ε, and averaged over the ($N \times N - 1$ values) for each delay. We have shown both analytical (equations (11) and (15)) and numerically computed results. Numerical results yield a very close fit to analytical results. Average cross correlation is of course only one statistic of interest.[1] The lower mean cross correlation for the quartic set for small delay values is evident.

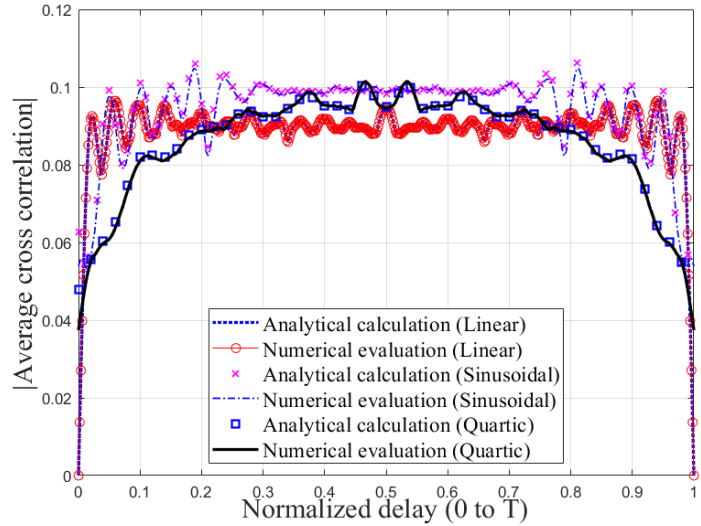

Fig. 5. Average values of analytical and numerically evaluated cross correlation versus delay for three different chirp signal sets, each with $N$= 25 users.

---

[1]An interactive graphical MATLAB application developed to find cross correlation numerically and analytically using (5) along with time, time-frequency domain waveform graphics, is available in: http://bit.ly/37KPJyL

Fig. 6 (a) to (c) show average correlation values for these three chirp types for three different values of the number of signals $N$. Insets in the figure show these correlations at two smaller delay ranges, $0.05T$ and $0.01T$, for illustration. We observe that beyond a certain small value of delay, the quartic nonlinear signals yield a smaller average correlation value for nearly the entire range of timing offset for the two smaller values of $N$, whereas the sinusoidal signals have approximately the same mean correlations as the linear case. Note that correlation plots are symmetric around $0.5T$ as Fig. 5 depicts, therefore only delays up to this value are shown in Fig. 6. Even for the largest value of $N$, the quartic signal set has lower mean correlations at delays above some very small value (~$0.005T$) up to a substantial delay value of approximately $0.1T$; these results illustrate the quartic set's suitability for QS operation.

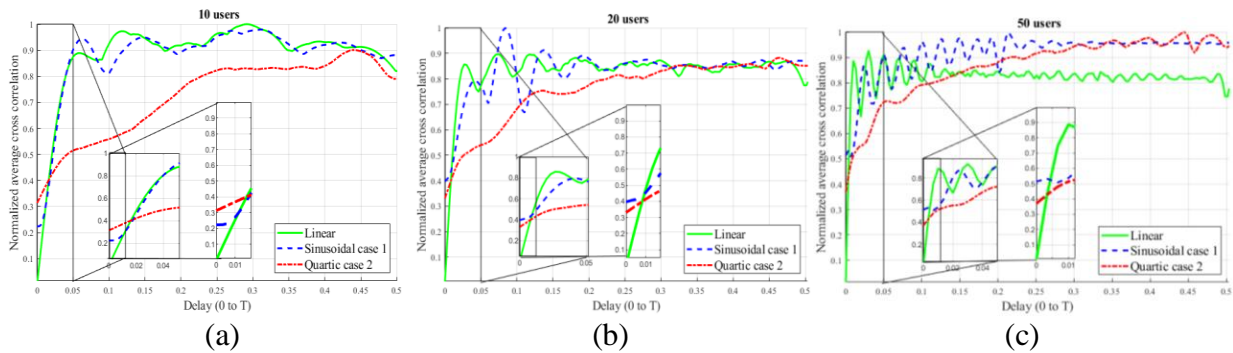

Fig. 6. Normalized average cross correlation versus delay for linear, sinusoidal, and quartic cases for (a) $N=10$, (b) $N=20$ and (c) $N=50$ signals.

For a more complete representation of the cross correlation distributions for these chirp types, we provide histograms of all cross correlation values for all offsets for our three signal sets in Fig. 7 (a) to (c). The histograms show that the largest correlation values, which cause the most severe MAI, are less likely for the nonlinear cases than the linear set.

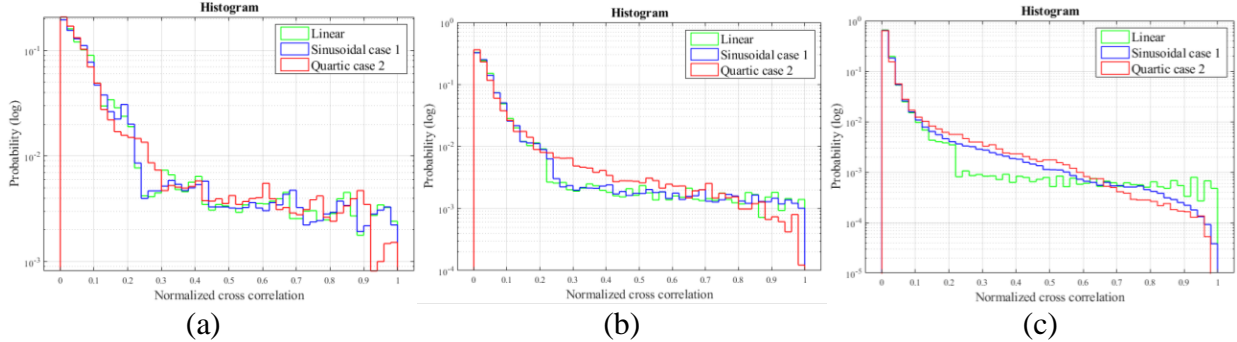

Fig. 7. Probability of occurrence of normalized cross-correlations for linear, sinusoidal and quartic chirps, for (a) *N*=10, (b) *N*=20 and (c) *N*=50 signals.

## V. Analytical Performance Evaluation

For a multiuser *M*-ary orthogonal linear chirp spread spectrum system the *k*'th user's transmitted baseband signal is,

$$s_k(t) = A_k \sum_{n=0}^{J-1} e^{\frac{j\pi N}{T^2}\left((t-nT)+\frac{kT}{N}+vT\right)^2} p(t-nT) \tag{16}$$

where $A_k$ is the signal amplitude, $k \in \{0,1,\dots,N-1\}$ is user index, $v$ is an equiprobable *M*-ary symbol $v \in \{0,1,\dots,M-1\}$, *T* is the symbol duration and function *p(t)* is the unit rectangular pulse equal to one for $0 < t < T$ and zero otherwise. This equation expresses transmission of a block of *J* symbols.

Fig. 8 illustrates the system block diagram. In the transmitter, for each user's data, a block of *b* bits is translated to one of $M=2^b$ symbols. Each symbol is mapped to a specific one of *M* sub-bands, and within each sub-band, a set of *N* chirp waveforms is used to accommodate the *N* users. Each sub-band has bandwidth *2N/T*, so the entire system bandwidth is *2NM/T* and the spectral efficiency of a fully loaded system is $log_2(M)/(2M)$ bps/Hz. In this paper we restrict our analysis to the binary CSS (BCSS) case. We also conduct the derivation beginning with linear chirps, but as shown in the Appendix, our actual result is applicable to any nonlinear chirp set as well, with the key requirement being that we have the cross correlation expressions to use within the BER formula.

We first assume an additive white Gaussian noise (AWGN) channel, and hence can consider detection during a single symbol interval. Performance is evaluated for user $k$ with $N$ user signals present, as illustrated in Fig. 8. After coherent downconversion, the baseband signal including noise at user $k$'s receiver can be written as,

$$y_k(t) = \sum_{i=0}^{N-1} s_i(t - \varepsilon_i) + w_k(t) \tag{17}$$

where $w_k(t)$ is complex noise $\left(n_{Ik}(t) + jn_{Qk}(t)\right)$ stationary and Gaussian with zero mean, $s_i(t - \varepsilon_i)$ is the signal of user $i$ at the receiver of user $k$, and $\varepsilon_k=0$. Note that in quasi-synchronous mode the delays ($\varepsilon$) can take any values; in general they are arbitrary and modeled as random, but we assume that they are constant for at least $J$ symbols.

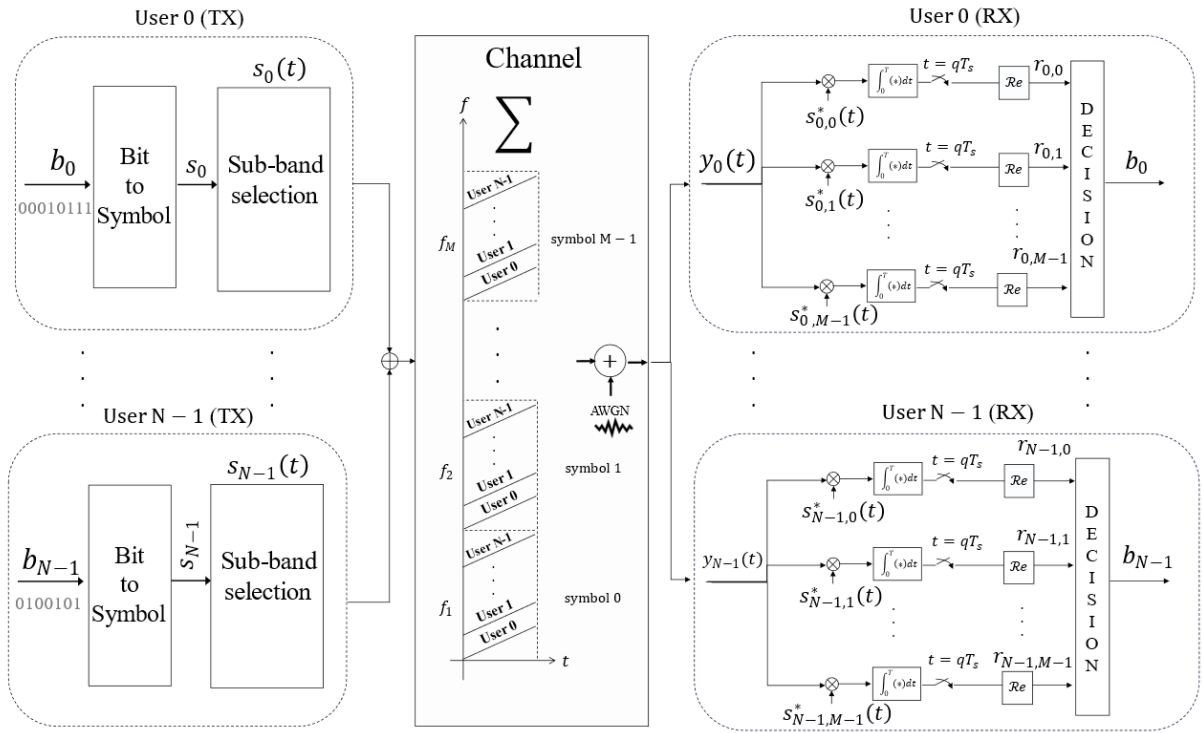

Fig. 8. Multiuser M-CSS complex baseband system block diagram

At the receiver, matched filters convolve the received signal with a bank of time-reversed versions of the transmitted chirps. An alternative heterodyne detector (correlator) can also be used instead of matched filter detectors, as explained in [26]. Decision circuits complete the receiver symbol detection.

Assuming user $k$ sends symbol "0", the BCSS decision block inputs for the two correlator branches can be written as,

$$\begin{cases} r_{k,0} = \int_0^{T_s} \left[ \sum_{i=0}^{N-1} s_i(t - \varepsilon_i) + w_k(t) \right] s_{k0}^*(t)\, dt \\ r_{k,1} = \int_0^{T_s} \left[ \sum_{i=0}^{N-1} s_i(t - \varepsilon_i) + w_k(t) \right] s_{k1}^*(t)\, dt \end{cases} \tag{18}$$

Analysis is analogous for the transmission of "1." Based upon the transmitted symbols, and using the expression derived for cross-correlations, we find that the correlator outputs are Gaussian with variance $\frac{N_0 A_k^2 T_s}{2}$ (the noise variance), and means dependent on the data symbols and cross-correlation values. Hence, we can find the bit error probability in terms of the well-known $Q$-function, the tail integral of the zero-mean, unit-variance Gaussian probability density function. For more details on the derivation, see the Appendix. For $N$ users in a binary ($M=2$) CSS system, the resulting BER for user $k$ can be expressed as,

$$P_{b,k} = \frac{1}{2^{N-1}} \sum_{\xi=0}^{2^{N-1}-1} Q\left( \sqrt{\frac{(1 + \boldsymbol{\rho}_k^T \mathbf{b}_\xi)^2 E_{s_k}}{N_0}} \right) \tag{19}$$

where $\mathbf{b}_\xi$ is a vector of size $(N-1) \times 1$, with elements in the set $\{-1, 1\}$, defined as,

$$\mathbf{b}_\xi = \begin{bmatrix} (-1)^{a(\xi,0)} \\ (-1)^{a(\xi,1)} \\ \vdots \\ (-1)^{a(\xi,N-2)} \end{bmatrix} \tag{20}$$

with $a(\xi, i) \in \{0,1\}\}$ the $i$th coefficient in the binary expansion for decimal number $\xi$, i.e,

$$\xi = \sum_{i=0}^{N-1} a(\xi, i) 2^i \qquad (21)$$

and $\boldsymbol{\rho}$ is the cross correlation vector of dimension $(N-1) \times 1$, with superscript $T$ denoting transpose. This vector is

$$\boldsymbol{\rho}_k = \boldsymbol{\rho}_{km}\{\boldsymbol{\rho}_{kk}\} = \begin{bmatrix} \rho_{k0} \\ \rho_{k1} \\ \vdots \\ \rho_{k,N-1} \end{bmatrix}, \qquad (22)$$

i.e., vector $\boldsymbol{\rho}_k$ includes all cross-correlation values $\rho_{km}$ except $\rho_{kk}$. For unequal received signal energies, the cross-correlation vector can be written as,

$$\boldsymbol{\rho}_k = \left[ \sqrt{\frac{E_{s_0}}{E_{s_k}}} \rho_{k0} \quad \sqrt{\frac{E_{s_1}}{E_{s_k}}} \rho_{k1} \quad \cdots \quad \sqrt{\frac{E_{s_{N-1}}}{E_{s_k}}} \rho_{k,N-1} \right]^T. \qquad (23)$$

As an example for evaluating (20) and (21), the **b** vectors for $N=4$ are as follows:

$$\mathbf{b}_0 = \begin{bmatrix} 1 \\ 1 \\ 1 \end{bmatrix}, \mathbf{b}_1 = \begin{bmatrix} -1 \\ 1 \\ 1 \end{bmatrix}, \mathbf{b}_2 = \begin{bmatrix} 1 \\ -1 \\ 1 \end{bmatrix}, \mathbf{b}_3 = \begin{bmatrix} -1 \\ -1 \\ 1 \end{bmatrix},$$

$$\mathbf{b}_4 = \begin{bmatrix} 1 \\ 1 \\ -1 \end{bmatrix}, \mathbf{b}_5 = \begin{bmatrix} -1 \\ 1 \\ -1 \end{bmatrix}, \mathbf{b}_6 = \begin{bmatrix} 1 \\ -1 \\ -1 \end{bmatrix}, \mathbf{b}_7 = \begin{bmatrix} -1 \\ -1 \\ -1 \end{bmatrix}, \qquad (24)$$

A comparison of simulation and analytical results for a binary linear chirp system with two users and different fixed delay values is presented in Fig. 9, showing essentially perfect agreement. An example for unequal energies is also included. Fig. 10 shows BER vs. SNR for a fixed delay value of $0.1T$ for different numbers of users. Here we assumed the desired user is synchronized and all other user signals have delay $\epsilon$. We again include an example for unequal energies. In both Fig. 9 and 10 we see excellent agreement between analytical and simulation results. Note that each user's performance is related to their position within the set and the delays of other user signals.

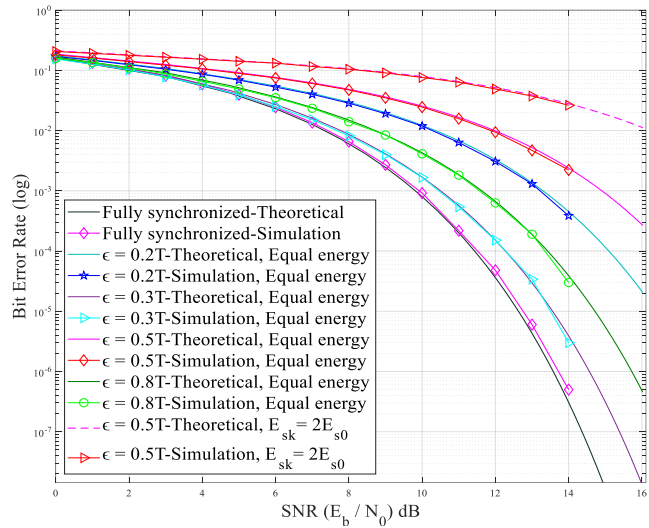

Fig. 9. BER vs. SNR for analysis and simulation for two users (*N*=2) in a linear BCSS system, for several values of delay and for both equal and unequal bit energies.

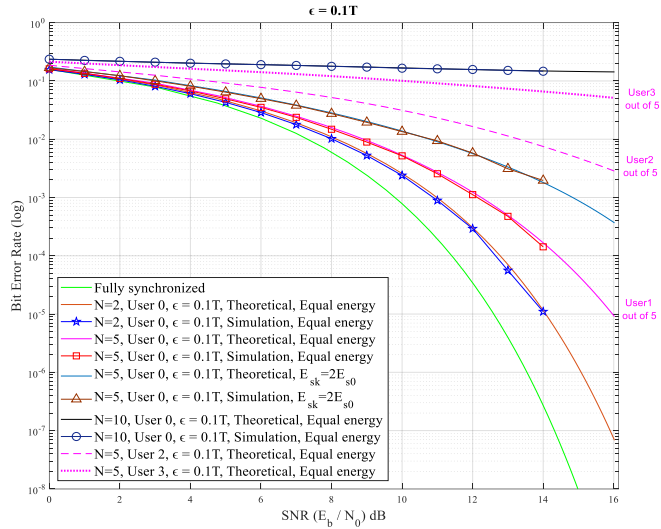

Fig. 10.     BER vs. SNR for linear BCSS comparing simulation versus theoretical results for different numbers of users with fixed timing offset and equal/unequal energy.

## VI. BER Performance Results

As previously noted, there are multiple ways to modulate chirps with data: mapping *M*-ary symbols to *M* of the *N* chirps in the set, using chirps of the opposite slope (e.g., "downchirps" as well as "upchirps"), on-off signaling, and even using different starting/stopping frequencies. This latter method is used in the LoRa technology [6], where with a linear chirp frequency $f_{chirp}$ in the range [$f_{min}$, $f_{max}$], two different symbols can be represented during a symbol interval by either (a) a sweep from $f_{min}$ to $f_{max}$, or by (b) a sweep from $f_m$ to $f_{max}$ immediately followed by sweep from $f_{min}$ to $f_m$, with the second symbol's start frequency $f_m$ in the range $f_{min} < f_m < f_{max}$.

As also noted, each user signal's delay is assumed perfectly known at its receiver. Delay tracking using coherent delay-locked loops (DLLs), similar to previous efforts for CDMA systems [27], can address this, but this is out of the scope of this paper. Our simulated performance results assume perfect delay estimation for each single-user receiver. Specifically, we model each delay as a zero-mean Gaussian random variable with standard deviation ($\sigma$).

*A. Quasi-synchronous vs. Fully Synchronized*

Fig. 11 (a) to (c) depict simulated bit error ratio performance versus bit energy to noise density ratio ($E_b/N_0$) for fully loaded linear and nonlinear designs for both synchronized and quasi-synchronous conditions for $N = 10, 20,$ and $50$. For these results, the zero-mean Gaussian random delays have standard deviation of $0.01T$ and $0.1T$. Fig. 11 (a) shows system performance in a perfectly synchronized system. The first thing to observe is that the nonlinear chirp signals are not orthogonal. Hence their performance degrades as the number of signals and MAI increase, particularly for the sinusoidal nonlinear case. However, as we can see in Fig. 11 (b), a *very* small set of random delays with $\sigma = 0.01T$ significantly degrades the performance of the linear chirps, whereas the degradation of the quartic nonlinear case is moderate. For the largest value of $\sigma$ in Fig. 11 (c), the nonlinear quartic case 2 is superior to the other sets of waveforms for any system loading. To show that our BER equation (19) can be used for any TF waveform shape, we also show analytical sinusoidal and quartic chirp performance in Fig. 11 (c).

Note that for actual random delays, we model the ε's as (Gaussian) random variables. To analytically assess performance in this case we would consider our cross correlations ((11) or (15)) to be *conditioned* on delay ε. Average BER would then be expressed as the integral of (19) multiplied by the Gaussian probability density function for ε. In general, the resulting complicated expression is not integrable in closed form; we leave exploration of this for future work.

Since the sinusoidal chirps do not extend the TF plane area coverage by much over the linear chirps, the sinusoidal chirps only slightly outperforms the linear case in Fig. 11 (b) and (c) in QS conditions. Synchronization on the order of $\sigma < 0.01T$ is very close to perfect, but the $0.1T$ value is more practical, particularly for mobile platforms.

Also worth study is performance of partially-loaded systems. As Fig. 12 depicts, the quartic CSS performance gain over the linear chirps in QS conditions will increase even more with the use of fewer signals ($K < N$=40) when the $K$ signals are selected to be maximally and equally spaced in the TF plane. This behavior has been observed for any arbitrary value of $N$.

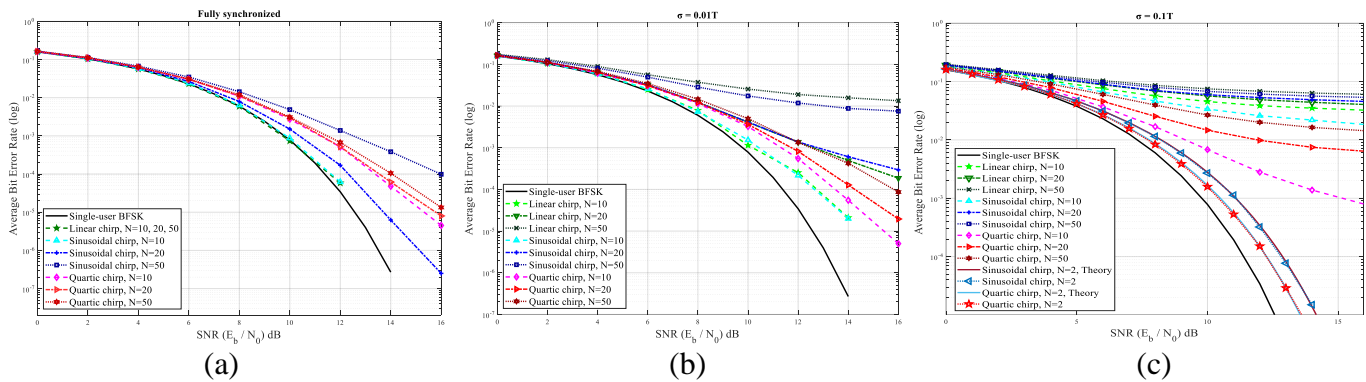

Fig. 11. Simulated multiuser binary chirp spread spectrum BER vs. $E_b/N_0$ for fully loaded systems for, (a) synchronized CSS system for $N$=10, $N$=20 and $N$=50, (b) quasi-synchronized CSS system with $\sigma = 0.01T$ for $N$=10, $N$=20 and $N$=50, and (c) quasi-synchronized CSS system with $\sigma = 0.1T$ for $N$=10, $N$=20 and $N$=50, including two analytical results.

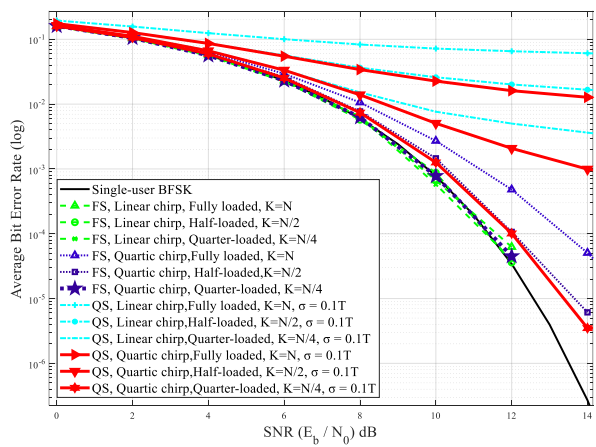

Fig. 12. Half- and quarter-loaded system simulated BER vs. $E_b/N_0$ for linear and quartic chirp signals in fully synchronized and QS conditions with $\sigma$=0.1$T$, and $N$=40.

## B. Proposed Nonlinear Chirps Versus Chirps from Literature

To further illustrate performance gains of our nonlinear chirp designs, we compare the performance of our nonlinear chirps with other chirp waveforms in the literature [7], [9] and [10] in quasi-synchronous conditions. This includes the amplitude-varying linear chirp, the quadratic, exponential, and hyperbolic sinusoidal. Fig. 13 shows BER vs. $E_b/N_0$ for $\sigma = 0.1T$ and $\sigma = 0.01T$.

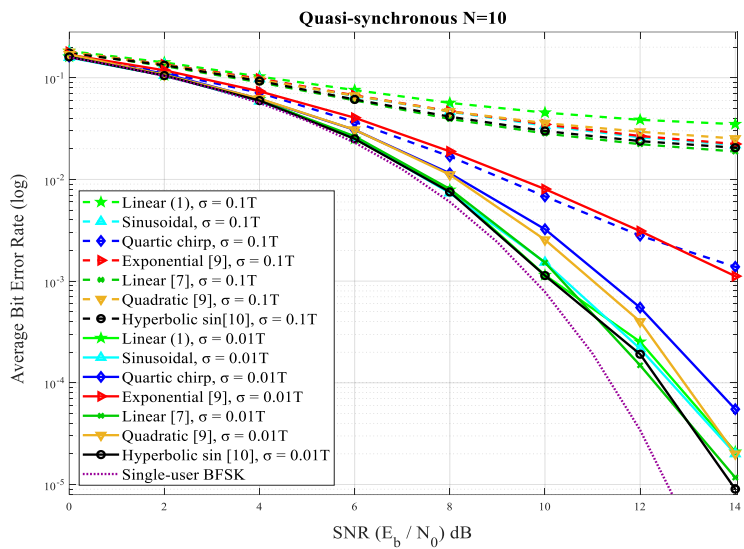

Fig. 13. Simulated multiuser binary chirp spread spectrum BER vs. $E_b/N_0$ for fully loaded systems for, (a) synchronized CSS system for $N=10$, $N=20$ and $N=50$, (b) quasi-synchronized CSS system with $\sigma = 0.01T$ for $N=10$, $N=20$ and $N=50$, and (c) quasi-synchronized CSS system with $\sigma = 0.1T$ for $N=10$, $N=20$ and $N=50$, including two analytical results.

Our quartic nonlinear set outperforms the other chirp waveforms for the practical case of $\sigma = 0.1T$. For the smaller value of $\sigma = 0.01T$, performance of all sets is very close except for the poorest-performing exponential case of [9]. Note that this plot is for a fully loaded system of 10 users. Waveforms from [7], [9] and [10] have different bandwidths for each user signal but the same total bandwidth was set to be identical for all selected waveforms. Moreover, all the other chirp signals in these references have amplitude variation, yielding a larger peak-to-average power ratio, whereas our waveforms have a constant envelope.

## C. Performance in an Empirical Air-Ground Channel

To finish description of our performance results, we simulated CSS performance over a dispersive air-ground channel. The channel models are based on empirical air to ground measurement results sponsored by

NASA, reported in [28] – [31]. Table I lists channel parameters for two locations: suburban Palmdale, CA, and the near urban setting for Cleveland, OH.

TABLE I. AIR-GROUND CHANNEL PARAMETERS [30].

| Parameters | Suburban Palmdale, CA | Near urban Cleveland, OH |
|---|---|---|
| *Mean RMS delay spread* | 53.78 ns | 16.6 ns |
| *Maximum RMS delay spread* | 1.2541 µs | 70.13 ns |
| *Frequency* | 5.06 GHz | 5.06 GHz |
| *Sounding bandwidth* | 50 MHZ | 50 MHZ |
| *Altitude* | 850 m | 850 m |

As can be seen, RMS delay spreads are larger for suburban Palmdale than for the near urban Cleveland channel. Since we employ no equalization or multipath mitigation in these initial results, we expect poorer performance in the suburban case.

The links for the example AG channels are a set of air to ground links emulating a multipoint to point air to ground system with total data rate of 100 kbits/s and total bandwidth of 400 kHz. There are $N$=10 users, each transmitting at 10 kbps over this bandwidth, the value of which is comparable to that proposed for other AG systems [32]. Transmissions from aircraft are received at the ground station quasi-synchronously, with zero-mean Gaussian distributed timing offsets with σ=0.1$T$, with each AG signal encountering its own unique channel.

Fig. 14 shows BER performance of the CSS signals over these realistic AG channels. For this relatively small bandwidth, the channel fading is essentially flat, except for the largest values of delay spread, which occur with low probability.

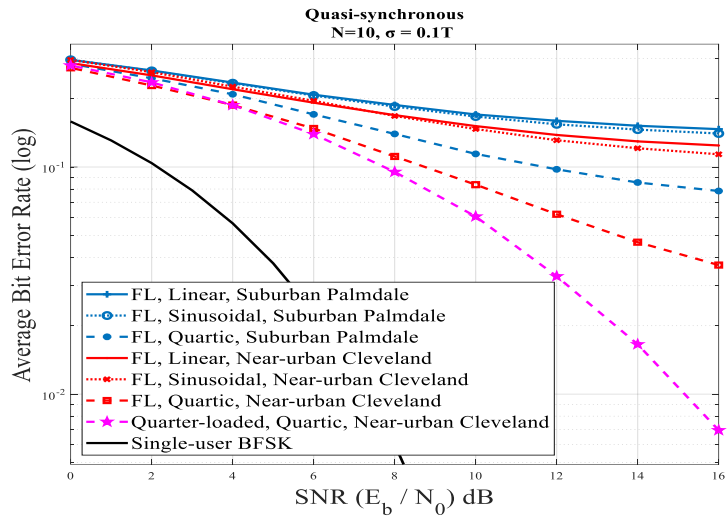

Fig. 14. Simulated BER vs. $E_b/N_0$ for CSS signals over simulated air-ground channels based on models in [30].

As expected based on delay spreads, results are better for near urban Cleveland than for suburban Palmdale. Performance of the quarter-loaded quartic system again illustrates the substantial effect of MAI in this system, but once again, our non-linear designs outperform the traditional linear chirps.

## VII. Conclusion

In this paper, we investigated multi user chirp spread spectrum system performance in quasi-synchronous conditions, for the classic linear chirp, several existing nonlinear chirp designs in the literature, and two new nonlinear chirps. We derived a closed-form expression for the cross correlation for the linear chirps and a closed-form approximation for nonlinear chirp cross correlations. From these we developed an expression for the error probability for any TF shape waveform for binary multi-user chirp spread spectrum in quasi-synchronous conditions. We validated our analysis via numerical and simulation results, and provided example correlation statistics for the linear and our new nonlinear chirps. The linear chirps are generally best in perfectly synchronized cases, but we showed that since our nonlinear cases use more "time-frequency space," they can outperform nearly all other chirp designs we have evaluated, for a range of assumed Gaussian-distributed timing offsets. The performance of our new designs is particularly superior in non-fully-loaded systems. Our new quartic nonlinear chirp design performs best. We also illustrated performance improvements of our new designs over a realistic dispersive air-ground channel. For future research, we will investigate the effects of Doppler shifts, and non-coherent detection for even more practical conditions.

## APPENDIX

Here we derive the BER expression for any user in the BCSS system, in the presence of up to *N*-1 asynchronous other-user signals. The block diagram for the *M*-ary CSS system was presented in Fig. 8 and the transmitted baseband waveform expression for user *k*'s signal is (16).

Without loss of generality, we derive the BER for user *k*=0, in the presence of a single interfering (asynchronous) user signal, user *m*=1. We then show that it is straightforward to generalize to an arbitrary number of interfering users, and that the expression holds for any selected user. We also analyze assuming that user 0 transmits a symbol $v$=0; with equally-likely data symbols, the derivation and results are identical for transmission of the symbol $v$=1. For our AWGN channel we can analyze detection of a single symbol to determine BER, and we assume transmission of the first bit, from time *t*=0 to *T*.

In our system, the 0 symbols all lie within the same sub-band, whereas all the 1 symbols lie in the adjacent sub-band; for the AWGN channel there is no "inter-sub-band interference." We assume that user 0's receiver is synchronized to its transmission, and the receiver correlates the received signal with the complex conjugate of the user 0 transmitted signal for both possible symbols 0 and 1. The bit decision is made by selecting the largest correlator output. Interfering user *m*=1's transmission is equally likely to be either a 0 or a 1, so we account for both possibilities. The interfering user *m*=1 signal is delayed by ε relative to user 0. The cross correlation between these two signals in the same sub-band ("0" sent) is given in (11) and (15). User *k*'s (=0) correlator outputs are given by,

$$\begin{matrix} r_{0,0} = \int_0^T [s_{00}(t) + s_{10}(t-\varepsilon) + w(t)] \, s_{00}^*(t) \, dt \\ r_{0,1} = \int_0^T [s_{00}(t) + s_{10}(t-\varepsilon) + w(t)] s_{01}^*(t) \, dt \end{matrix} \right\} \text{User1 sends "0"} \\ \begin{matrix} r_{0,0} = \int_0^T [s_{00}(t) + s_{11}(t-\varepsilon) + w(t)] \, s_{00}^*(t) \, dt \\ r_{0,1} = \int_0^T [s_{00}(t) + s_{11}(t-\epsilon) + w(t)] s_{01}^*(t) \, dt \end{matrix} \right\} \text{User1 sends "1"} \quad (A1)$$

where $r_{kv}$ denotes the user $k$ correlator output for symbol $v$, $s_{kv}(t)$ denotes user $k$'s signal for symbol $v$, and $w(t)$ is the AWGN. By expanding (A1), and using the definition of cross correlation we can write,

$$\left.\begin{aligned} r_{0,0} &= 2E_{s,0} + 2\rho_{01}(\epsilon)\sqrt{E_{s,0}E_{s,1}} + \int_0^T w(t)s_{00}^*(t)\,dt \\ r_{0,1} &= \int_0^T w(t)s_{01}^*(t)\,dt \end{aligned}\right\} \begin{aligned} &\text{User1} \\ &\text{sends"0"} \end{aligned}$$

$$\left.\begin{aligned} r_{0,0} &= 2E_{s,0} + \int_0^T w(t)s_{00}^*(t)\,dt \\ r_{0,1} &= 2\rho_{01}(\epsilon)\sqrt{E_{s,0}E_{s,1}} + \int_0^T w(t)s_{01}^*(t)\,dt \end{aligned}\right\} \begin{aligned} &\text{User1} \\ &\text{sends"0"} \end{aligned} \quad (A2)$$

Variable $r_{0,0}$ consists of the desired signal, MAI, and noise components when user 1 sends a 0, and desired signal and noise terms when user 1 sends a 1, whereas $r_{0,1}$ consists of noise only when user 1 sends a 0, and noise plus an MAI term when user 1 sends 1. The integrals involving the noise terms are zero-mean Gaussian variables with variance $N_0 E_{s,0}$.

In the binary case the decision can be cast as comparing the Gaussian variable $r_{0,0}-r_{0,1}$ with a threshold, and using the well-known tail integral of the zero-mean, unit variance Gaussian density function—the $Q$-function—we can express the error probability as,

$$P_{b,0} = \frac{1}{2}Q\left(\frac{E_{s,0} + \rho_{01}(\epsilon)\sqrt{E_{s,0}E_{s,1}}}{\sqrt{N_0 E_{s,0}}}\right) + \frac{1}{2}Q\left(\frac{E_{s,0} - \rho_{01}(\epsilon)\sqrt{E_{s,0}E_{s,1}}}{\sqrt{N_0 E_{s,0}}}\right)$$

$$= 0.5Q\left(\sqrt{\frac{\left(1 + \sqrt{\frac{E_{s,1}}{E_{s,0}}}\rho_{01}(\epsilon)\right)^2 E_{s,0}}{N_0}}\right) + 0.5Q\left(\sqrt{\frac{\left(1 - \sqrt{\frac{E_{s,1}}{E_{s,0}}}\rho_{01}(\epsilon)\right)^2 E_{s,0}}{N_0}}\right) \quad (A3)$$

As required, when synchronized ($\rho_{01} = 0$), the result reduces to the well-known result for coherent binary FSK.

For the performance with three users we have two additional $Q$-function terms, with each of the four terms multiplied by ¼ to account for the four equiprobable possibilities for the two interfering user symbol values:

$$P_{b,0} = \frac{1}{4}Q\left(\sqrt{\frac{\left(1+\sqrt{\frac{E_{s,1}}{E_{s,0}}}\rho_{01}(\epsilon)+\sqrt{\frac{E_{s,2}}{E_{s,0}}}\rho_{02}(\epsilon)\right)^2 E_{s,0}}{N_0}}\right) + \frac{1}{4}Q\left(\sqrt{\frac{\left(1-\sqrt{\frac{E_{s,1}}{E_{s,0}}}\rho_{01}(\epsilon)+\sqrt{\frac{E_{s,2}}{E_{s,0}}}\rho_{02}(\epsilon)\right)^2 E_{s,0}}{N_0}}\right) +$$

$$\frac{1}{4}Q\left(\sqrt{\frac{\left(1+\sqrt{\frac{E_{s,1}}{E_{s,0}}}\rho_{01}(\epsilon)-\sqrt{\frac{E_{s,2}}{E_{s,0}}}\rho_{02}(\epsilon)\right)^2 E_{s,0}}{N_0}}\right) + \frac{1}{4}Q\left(\sqrt{\frac{\left(1-\sqrt{\frac{E_{s,1}}{E_{s,0}}}\rho_{01}(\epsilon)-\sqrt{\frac{E_{s,2}}{E_{s,0}}}\rho_{02}(\epsilon)\right)^2 E_s}{N_0}}\right)$$

(A4)

By continuing this process of including additional asynchronous users, by induction we arrive at the final expression (19). In addition, since the BER equations account for the type of chirps only through the cross correlations, we also deduce that they pertain to *any* chirp type—linear or nonlinear—as long as we have the expressions for cross correlation to use within these BER equations. This claim is validated via our results in Section V.

## ACKNOWLEDGMENT

The authors thank Dr. H. Jamal of the University of South Carolina for development of the AG channel routines.

## REFERENCES

[1] S. D. Blunt and E. L. Mokole, "Overview of radar waveform diversity," *IEEE Aerospace and Electronic Systems Magazine*, vol. 31, no. 11, pp. 2-42, November 2016.

[2] *IEEE Wireless Medium Access Control (MAC) and Physical Layer (PHY) Specifications for Low-Rate Wireless Personal Area Networks (WPANs),* IEEE 802.15.4a, 2007.

[3] J. A. Catipovic, "Performance limitations in underwater acoustic telemetry," *IEEE Journal of Oceanic Engineering*, vol. 15, no. 3, pp. 205-216, July 1990.

[4] C. He, J. Huang, Q. Zhang and K. Lei, "Reliable Mobile Underwater Wireless Communication Using Wideband Chirp Signal," *2009 WRI International Conf. on Communications and Mobile Computing*, Yunnan, 2009, pp. 146-150

[5] N. Sornin, M. Luis, T. Eirich, T. Kramp, and O. Hersent, *LoRaWAN Specifications*, LoRa Alliance, San Ramon, CA, USA, 2015.

[6] O. Georgiou and U. Raza, "Low Power Wide Area Network Analysis: Can LoRa Scale?," *IEEE Wireless Comm. Letters*, vol. 6, no. 2, pp. 162-165, April 2017.

[7] H. Shen and A. Papandreou-Suppappola, "Diversity and channel estimation using time-varying signals and time-frequency techniques," *IEEE Trans. Signal Processing*, vol. 54, no. 9, pp. 3400-3413, Sept. 2006.

[8] X. Ouyang and J. Zhao, "Orthogonal Chirp Division Multiplexing," *IEEE Trans. Communications*, vol. 64, no. 9, pp. 3946-3957, Sept. 2016.

[9] M. A. Khan, R. K. Rao, and X. Wang, "Performance of quadratic and exponential multiuser chirp spread spectrum communication systems," *International Symp. on Performance Evaluation of Computer and Telecommunication Systems (SPECTS)*, Toronto, ON, 2013, pp. 58-63.

[10] M. A. Khan, R. K. Rao, and X. Wang, "Non-linear trigonometric and hyperbolic chirps in multiuser spread spectrum communication systems," *IEEE 9th International Conference on Emerging Technologies (ICET)*, Islamabad, Pakistan, 2013.

[11] S-M Tseng and M. R. Bell, "Asynchronous multicarrier DS-CDMA using mutually orthogonal complementary sets of sequences," *IEEE Trans. Communications*, vol. 48, no. 1, pp. 53-59, Jan. 2000.

[12] Z. Gu, S. Xie and S. Rahardja, "Performance analysis for DS-CDMA systems with UCHT signature sequences over fading channels," *2004 IEEE 59th Vehicular Technology Conference, VTC 2004-Spring*, Milan, IT, 2004.

[13] G. Gong, F. Huo and Y. Yang, "Large Zero Autocorrelation Zones of Golay Sequences and Their Applications," *IEEE Trans. Communications*, vol. 61, no. 9, pp. 3967-3979, September 2013.

[14] *The Industrial Reorganization Act: The communications industry*. (1973), Columbia Law Review, 73(3), 635-676.

[15] G. F. Gott and J. P. Newsome, "H.F. data transmission using chirp signals," *Proc.s of the Institution of Electrical Engineers*, vol. 118, no. 9, pp. 1162-1166, September 1971.

[16] Y. Ju and B. Barkat, "A new efficient chirp modulation technique for multi-user access communications systems," *IEEE International Conf. on Acoustics, Speech, and Signal Processing*, Montreal, QU, 2004.

[17] H. M. Ozaktas, Z. Zalevsky, M.A. Kutay *The Fractional Fourier Transform with Applications in Optics and Signal Processing*. John Wiley & Sons Ltd, 2000.

[18] R. Bomfin, M. Chafii and G. Fettweis, "Low-Complexity Iterative Receiver for Orthogonal Chirp Division Multiplexing," *2019 IEEE Wireless Communications and Networking Conference Workshop (WCNCW)*, Marrakech, Morocco, 2019, pp. 1-6.

[19] A. Sahin, N. Hosseini and D. W. Matolak, " DFT-spread-OFDM Based Chirp Transmission," *2020 IEEE 21th Wireless and Microwave Technology Conference (WAMICON)*, Clearwater Beach, FL, USA, 2020.

[20] A. Sahin, D. W. Matolak, N. Hosseini, "Methods for Reliable Chirp Transmissions and Multiplexing," U.S. Provisional Pat. Ser. No. 62913829, filed October 2019.

[21] N. Hosseini and D. W. Matolak, "Chirp Spread Spectrum Signaling for Future Air-Ground Communications," *MILCOM 2019 - 2019 IEEE Military Communications Conference (MILCOM)*, Norfolk, VA, USA, 2019, pp. 153-158.

[22] H. F. Talbot, "Facts Relating to Optical Science," *Dublin Philosophical Magazine and Journal of Science,* pp. 401-407, London 1836.

[23] J. Goodman, Introduction to Fourier Optics, 2nd ed. *McGrawHill*, 1966.

[24] J. T. Winthrop and C. R. Worthing, "Convolution formulation of Fresnel diffraction," *J. Opt. Soc. Am.*, vol. 56, pp. 588-591, 1966.

[25] S. Wolfram, "Wolfram Alpha intelligence computational integral engine," Internet: https://www.wolfram.com/, [June. 1, 2020].

[26] N. Hosseini and D. W. Matolak, "Wide band channel characterization for low altitude unmanned aerial system communication using software defined radios," *IEEE/AIAA Integrated Communications, Navigation, Surveillance Conference*, Herndon, VA, April 2018.

[27] M. Latva-aho and J. Lilleberg, "Delay trackers for multiuser CDMA receivers," *Proc. of ICUPC - 5th Int. Conf. on Universal Personal Communications*, Cambridge, MA, USA, 1996, pp. 326-330 vol.1.

[28] D. W. Matolak and R. Sun, "Air–Ground Channel Characterization for Unmanned Aircraft Systems—Part I: Methods, Measurements, and Models for Over-Water Settings," *IEEE Trans. Vehicular Tech.*, vol. 66, no. 1, pp. 26-44, Jan. 2017.

[29] R. Sun and D. W. Matolak, "Air–Ground Channel Characterization for Unmanned Aircraft Systems Part II: Hilly and Mountainous Settings," *IEEE Trans. Vehicular Tech.*, vol. 66, no. 3, pp. 1913-1925, March 2017.

[30] D. W. Matolak and R. Sun, "Air–Ground Channel Characterization for Unmanned Aircraft Systems—Part III: The Suburban and Near-Urban Environments," *IEEE Trans. Vehicular Tech.*, vol. 66, no. 8, pp. 6607-6618, Aug. 2017.

[31] R. Sun, D. W. Matolak and W. Rayess, "Air-Ground Channel Characterization for Unmanned Aircraft Systems—Part IV: Airframe Shadowing," *IEEE Trans. Vehicular Tech.*, vol. 66, no. 9, pp. 7643-7652, Sept. 2017.

[32] S. Brandes, et al. "Physical layer specification of the L-band Digital Aeronautical Communications System (L-DACS1)." *Integrated Communications, Navigation and Surveillance Conference (ICNS), IEEE*, pp. 1-12, Arlington VA, 13-15 May 2009.